\documentclass[fleqn,twoside]{article}
\usepackage{espcrc2}
\usepackage{epsfig}

\usepackage{graphicx}
\usepackage[figuresright]{rotating}

\newcommand{\AmS}{{\protect\the\textfont2
  A\kern-.1667em\lower.5ex\hbox{M}\kern-.125emS}}

\newcommand{\p}{\partial}
\newcommand{\beq}{\begin{equation}}
\newcommand{\eeq}{\end{equation}}

\title{Study of the 2-d $CP(N-1)$ models at $\theta=0$ and $\pi$}

\author{B.\ B.\ Beard\address[CBU]{Departments of Physics and Mechanical Engineering,
Christian Brothers University, \\ Memphis, TN 38104, U.S.A.}, 
M.\ Pepe\address[Bern]{Institute for Theoretical Physics, Bern University,
Sidlerstrasse 5, 3012 Bern, Switzerland}\thanks{Speaker at the Conference.},
S.\ Riederer\addressmark[Bern],
and U.-J.\ Wiese\addressmark[Bern]}
       
\begin{document}

\begin{abstract}
We present numerical results for 2-d $CP(N\! -\! 1)$ models at $\theta \!=\!0$ and $\pi$ obtained in
the D-theory formulation. In this formulation we construct an efficient cluster algorithm and we show
numerical evidence for a first order transition for $CP(N\! -\! 1\! \geq\! 2 )$ models at
$\theta = \pi$. By a finite size scaling analysis, we also discuss the equivalence in the
continuum limit of the D-theory formulation of the 2-d $CP(N-1)$ models and the usual lattice
definition.
\end{abstract} 

\maketitle

\section{Introduction}
$CP(N-1)$ models~\cite{DAd78} are defined on the manifold $S^{2N-1}/S^1$ which is the coset
space of the breaking of $SU(N)$ down to $U(N-1)$. The field $P(x)$
is parametrized by Hermitean $N\! \times\! N$ projector matrices, i.e. $P(x)^2 = P(x)$ and
$\mbox{Tr } P(x) = 1$. The action is given by
\beq
S[P] = \frac{1}{g^2} \int \;dx \;\mbox{Tr} [\p_\mu P(x) \p_\mu P(x)],
\eeq
and it is invariant under the global $\Omega \in SU(N)$ transformation
$P(x)'=\Omega P(x) \Omega^+$.
The 2-d $CP(N-1)$ models are interesting toy-models of 
4-d Yang-Mills theories. In fact, these two quantum field theories share several
important features, including asymptotic freedom, the dynamical generation of a
mass-gap and an instanton topological charge. In particular, this latter property leads to
a non-trivial vacuum angle $\theta$ and allows to consider $CP(N-1)$ models in order to
investigate $\theta$-vacuum effects. It has been conjectured that $CP(N-1)$ models have a
phase transition at $\theta\! = \! \pi$. For $CP(1)\! =\! O(3)$ this phase transition is known to
be second order with a vanishing mass-gap~\cite{Hal83,Aff86,Bie95}. For $N\! >\! 2$ the
transition is conjectured to be first order~\cite{Aff88,Sei84}. Studying these
nonperturbative problems is highly nontrivial and one can eventually 
address these questions only numerically. However, the numerical simulation of 
$CP(N\!-\!1)$ models is not straightforward, even at $\theta = 0$. In fact,
in contrast to $O(N)$ models where the Wolff cluster algorithm~\cite{Wol89} can be used,
no efficient cluster algorithm is available~\cite{Jan92,Car93}. Still a
rather efficient multi-grid algorithm was developed in~\cite{Has92}. 
At $\theta = \pi$ the numerical investigations become even more difficult due to a very
severe sign problem. In fact, in this case, the topological sectors with even 
and odd charges have opposite Boltzmann weights and their contributions almost completely
cancel in the partition function. This makes it exponentially hard to simulate the large
lattices necessary to reach reliable conclusions on the phase
structure. For this reason, efficient simulations at $\theta = \pi$ have been performed
only in the $CP(1) = O(3)$ case by a Wolff-type meron-cluster algorithm~\cite{Bie95}. 
Here we present results obtained with a method allowing to perform an unbiased and very
accurate numerical study of any 2-d $CP(N-1)$ model at $\theta = 0$ and $\pi$~\cite{BBB04}. Our
method is based on the {\em D-theory} formulation of a field theory~\cite{Cha97}. This is a
different formulation than Wilson's and consists in obtaining the continuous fields
from the {\em dimensional} reduction of {\em discrete} variables. It provides an alternative
lattice regularization of field theory but with the same continuum limit.
More details on the results presented here can be found in~\cite{BBB04}.

\section{$CP(N-1)$ models: D-theory formulation}
In this section we discuss 2-d $CP(N\!-\!1)$ models in the D-theory formulation. The
discrete variables we consider are $SU(N)$ quantum spins 
$T_x^a = \frac{1}{2} \lambda_x^a$, generators of the $SU(N)$ group 
$[T_x^a,T_y^b] =  i \delta_{xy} f_{abc} T_x^c$.  
The spins are located on the sites $x$ of a $L \times L'$ periodic 
square lattice with $L \gg L'$. Hence we have the geometry of a spin ladder 
consisting of $n = L'/a$ transversely coupled spin chains of length $L$. 
\begin{figure}[htb]
\vspace{-0.8cm}
\begin{center}
\epsfig{file=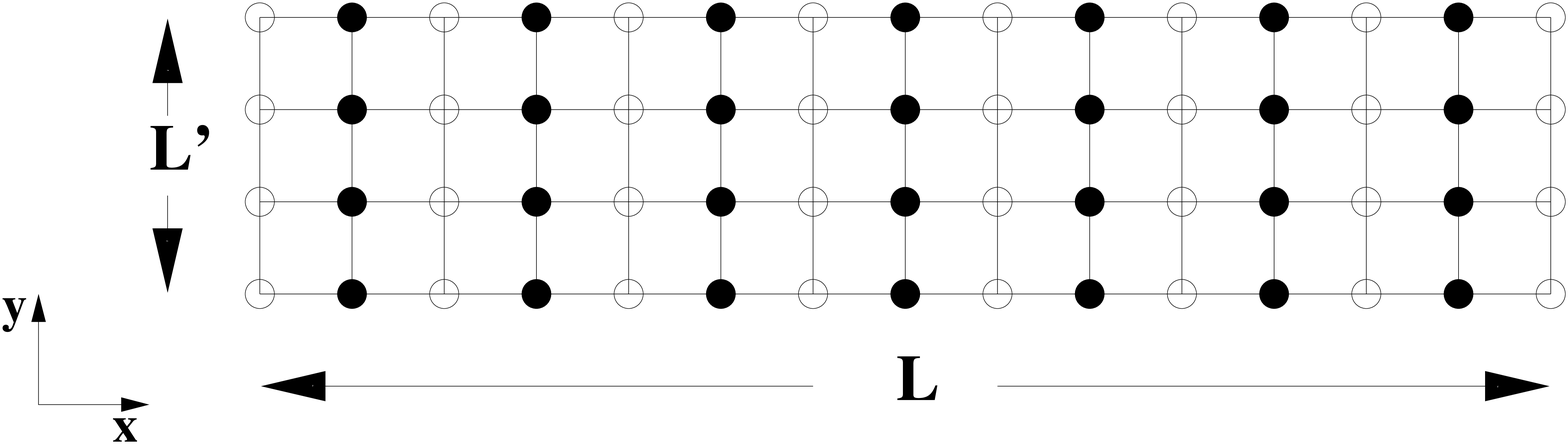,width=7.cm,angle=0}
\end{center}
\vspace{-1.4cm}
\caption{\it Spin ladder geometry: open circles form sublattice $A$,
filled circles belong to sublattice $B$.}
\vspace{-.9cm}
\end{figure}
We consider n.n.\ couplings: antiferromagnetic along the chains and 
ferromagnetic between different chains. The lattice naturally decomposes
into two sublattices $A$ and $B$. The spins in $A$ transform in the fundamental
representation $\{N\}$, while those belonging to $B$ are in the
conjugate one $\{\overline N\}$ and are described by $- T_x^{a *}$. The quantum spin
ladder Hamiltonian~is 
\begin{eqnarray}\label{hamilt}
H&=&- J \sum_{x \in A} [T_x^a T_{x+\hat 1}^{a *} + T_x^a T_{x+\hat 2}^a]
\nonumber \\
&-&J \sum_{x \in B} [T_x^{a *} T_{x+\hat 1}^a + T_x^{a *} T_{x+\hat 2}^{a *}],
\end{eqnarray}
where $J > 0$, and $\hat 1$ and $\hat 2$ are unit-vectors in the $x$-
and $y$-directions, respectively. The system has a global $SU(N)$ 
symmetry, i.e.\ $[H,T^a] = 0$, with the total spin 
$T^a = \sum_{x \in A} T_x^a - \sum_{x \in B} T_x^{a *}$,
satisfying the $SU(N)$ algebra $[T^a,T^b] = i f_{abc} T^c$.
The quantum partition function at the temperature $T=1/\beta$ is then given by 
$Z = \mbox{Tr exp} (-\beta H)$.

Using the coherent state technique of~\cite{Rea89} one finds that, at $T=0$ and for 
$L, L' \rightarrow \infty$, the $SU(N)$ symmetry breaks down to
$U(N-1)$. Thus there are massless Goldstone bosons described by fields 
$P\! \in\! SU(N)/U(N-1) = CP(N-1)$. When the $y$-direction is compactified to a finite
extent $L'$, the Coleman-Hohenberg-Mermin-Wagner theorem forbids massless 
excitations. Thus the Goldstone bosons pick up a nonperturbatively
generated mass-gap $m = 1/\xi$ and the correlation length $\xi$ is now finite.
For sufficiently many transversely coupled chains, the correlation length becomes
exponentially large, $\xi \gg L'$, and the system both undergoes dimensional reduction and
approaches the continuum limit. Using chiral perturbation theory, one finds that the
lowest-order terms in the 2-d Euclidean effective action are given by
\beq
S[P] = \int dx dt  \{ \frac{1}{g^2} \; \mbox{Tr}
[\p_\mu P \p_\mu P] - i \theta Q[P] \}.
\eeq
where $Q[P]$ is the topological charge
\beq
Q[P] = \frac{1}{2 \pi i} \int dx dt \; \epsilon _{\mu \nu}
\mbox{Tr}[P \p_\mu P \p_\nu P],
\eeq
and $\theta = n \pi$ is the vacuum angle. Hence, $\theta = 0$ for even $n$ and $\theta = \pi$
for odd $n$.   

\section{The numerical study}
One advantage of D-theory is that it is a formulation of a quantum field 
theory in terms of simple discrete degrees of freedom instead of the usual
continuum classical fields. In particular, the partition function $Z$ of the $SU(N)$
quantum spin ladder can be written using a basis of discrete $SU(N)$ spin states:  
$q \in \{u,d,s,...\}$ on sublattice $A$ and $\overline q \in \{\overline u,
\overline d, \overline s,...\}$ on sublattice $B$. These can be simulated with the 
loop-cluster algorithm~\cite{Eve93,Wie94}. The efficient cluster algorithm we have   
used in this study is a generalization to $SU(N)$ of the $SU(2)$ case~\cite{Wie94}.

The Hamiltonian (\ref{hamilt}) is the sum of 4 terms that we denote, respectively,
with $H_1,\ldots, H_4$: two of them ($H_2$, $H_4$) describe the ferromagnetic interaction
while the other two ($H_1$, $H_3$) the antiferromagnetic one. Using the Trotter
decomposition we write $Z$ as follows
\beq
Z=\lim_{M\rightarrow \infty} \mbox{Tr}\left[
\mbox{e} ^{-\epsilon H_1} \mbox{e} ^{-\epsilon H_2} 
\mbox{e} ^{-\epsilon H_3} \mbox{e} ^{-\epsilon H_4} 
\right] ^M
\eeq
where we have introduced $4M$ slices in the $\beta$-direction with spacing
$\epsilon=\beta J/M$. Inserting complete sets of states between the exponential factors, one
obtains that the only non-vanishing transfer matrix elements are given by
\begin{eqnarray}
&&\hspace{-.6cm} 
\langle q^i,q^i 
|\mbox{e} ^{-\epsilon H_2} | 
q^i,q^i \rangle =
\mbox{e} ^{\epsilon \frac{N-1}{2N}} \\
&&\hspace{-.6cm} 
\langle q^i,q^{j\neq i} 
|\mbox{e} ^{-\epsilon H_2} | 
q^i,q^{j\neq i} \rangle =
(\mbox{e} ^{\epsilon \frac{N-1}{2N}} + \mbox{e} ^{-\epsilon \frac{N+1}{2N}})/2
\nonumber \\
&&\hspace{-.6cm} 
\langle q^i,q^{j\neq i} 
|\mbox{e} ^{-\epsilon H_2} | 
q^{j\neq i}, q^i \rangle = 
(\mbox{e} ^{\epsilon \frac{N-1}{2N}} - \mbox{e} ^{-\epsilon \frac{N+1}{2N}})/2
\nonumber \\
&&\hspace{-.6cm} 
\langle q^i,\overline{q}^i 
|\mbox{e} ^{-\epsilon H_1} | 
q^i,\overline{q}^i \rangle =
(\mbox{e} ^{\frac{\epsilon N}{2}} +N-1)\, \mbox{e} ^{-\frac{\epsilon }{2N}}/N
\nonumber \\
&&\hspace{-.6cm} 
\langle q^i,\overline{q}^{j\neq i} 
|\mbox{e} ^{-\epsilon H_1} | 
q^i,\overline{q}^{j\neq i} \rangle = 
\mbox{e} ^{-\frac{\epsilon }{2N}}
\nonumber \\
&&\hspace{-.6cm} 
\langle q^i,\overline{q}^{j\neq i} 
|\mbox{e} ^{-\epsilon H_1} | 
q^{j\neq i}, \overline{q}^i \rangle = 
(\mbox{e} ^{\frac{\epsilon N}{2}} - 1)\, \mbox{e} ^{-\frac{\epsilon }{2N}}/N
\nonumber 
\end{eqnarray}
The transfer matrix elements for $H_3$ and $H_4$ can be obtained by substituting 
$H_1\! \rightarrow\! H_3$,
$H_2\! \rightarrow\! H_4$ and $q\! \leftrightarrow\! \overline{q}$. With these matrix elements at
hand, the construction of the quantum spin cluster and its updating proceeds 
like in the~$SU(2)$~case. One can also avoid the slicing in the $\beta$-direction
operating directly in continuous Euclidean time~\cite{Bea96}. 

We have used this cluster algorithm to investigate if $CP(N-1)$ models with
$N>2$ have a first order phase transition at $\theta = \pi$ where the charge 
conjugation symmetry $C$ is spontaneously broken. An order parameter for such a
phase transition would be the topological charge $Q[P]$ which indeed is $C$-odd:
$Q[^CP] = Q[P^*] =  - Q[P]$. 
However $Q$ is defined only in the framework of the target
continuum theory. In the discrete spin system we define as order parameter 
$Q[q,\overline q]$ the number of spin flips in a 
configuration. $Q[q,\overline q]$ gets a contribution +1 if a pair of n.n.\ spins along the
$x$-direction, $q_x \overline q_{x+\hat 1}$, flips to another state, 
$q_x' \overline q_{x+\hat 1}'$, at some moment in time. A spin flip from 
$\overline q_x q_{x+\hat 1}$ to $\overline q_x' q_{x+\hat 1}'$ instead contributes
$-1$. In the quantum spin ladder, charge conjugation corresponds to replacing each spin
state $q_x$ by  $q_{x+\hat 1}$ and, indeed, $Q[q,\overline q]$ changes sign under this
operation while the action remains invariant.

\begin{figure}[tb]\label{CP3}
\epsfig{file=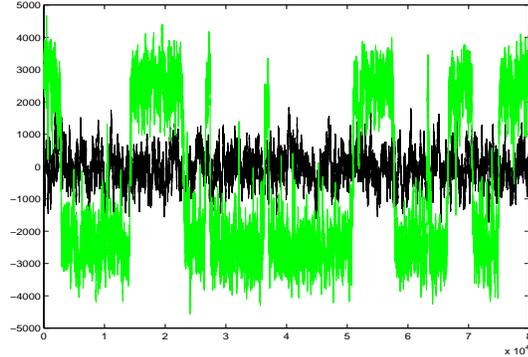,width=7.cm,height=4.8cm,angle=0}
\vspace{-1.1cm}
\caption{\it Monte Carlo time histories of 
$Q[q,\overline q]$ for the $CP(3)$ model at $\theta = 0$ and $\pi$.}
\vspace{-1.cm}
\end{figure}

We have simulated $SU(N)$ quantum spin ladders for
$N = 3,4$, and 5, with $n = L'/a \in \{2,3,...,7\}$. Our simulations confirm the existence
of a first order phase transition with spontaneous $C$-breaking at $\theta = \pi$ for all 
$N>2$. As expected, there is no phase transition at $\theta = 0$. Figure \ref{CP3}
shows Monte Carlo time histories of $Q[q,\overline q]$ for $SU(4)$ spin ladders
with $n = 3$ and 4. For $n = 3$ one observes two coexisting phases with spontaneous 
$C$-breaking, while for $n = 4$ there is only one $C$-symmetric phase.
Similar results have been obtained in the other cases studied.

Finally, we have verified explicitly that the $CP(N-1)$ models in Wilson's and in the
D-theory formulation have the same continuum limit. In order to do this we have compared a 
physical quantity  -- namely the universal finite-size scaling function 
$F(\xi(L)/L) = \xi(2 L)/\xi(L)$ -- measured in both frameworks. In~\cite{BBB04} we show the
excellent agreement between the results obtained in these two formulations.

\end{document}